\documentclass[prb,aps,preprint]{revtex4}
\newcommand{\beq}{\begin{equation}}
\newcommand{\eeq}{\end{equation}}
\usepackage{graphics}
\usepackage{epsf}
\usepackage{epsfig}
\usepackage{amsmath}
\usepackage{amsfonts}
\usepackage{amssymb}
\usepackage{graphicx}
\usepackage{epsfig}
\begin{document}

\title{Origin of the unconventional magnetoresistance in Sr$_2$FeMoO$_6$}

\author{Sugata Ray,$^{1,2,\star}$ %
Srimanta Middey,$^{2}$ %
Somnath Jana,$^{2}$     %
A. Banerjee,$^{3}$   %
P. Sanyal,$^{4}$ %
Rajeev Rawat,$^{3}$ %
Luca Gregoratti,$^5$ %
and D. D. Sarma$^{6}$}

\affiliation{$^1$Department of Materials Science, Indian Association for the Cultivation of Science, Jadavpur, Kolkata 700 032, India\\
$^2$Centre for Advanced Materials, Indian Association for the Cultivation of Science, Jadavpur, Kolkata 700 032, India\\
$^3$UGC-DAE CSR, University Campus, Khandawa Road, Indore 451 027, India\\
$^4$S. N. Bose National Centre for Basic Sciences, Sector-III, Block - JD, Kolkata 700 098, India\\
$^5$Sincrotrone Trieste, Area Science Park, 34012 BasoVizza-Trieste, Italy\\
$^6$Solid State and Structural Chemistry Unit, Indian Institute of Science, Bangalore 560 012, India}

%\newpage

\begin{abstract}
{The unusual magnetoresistance ($MR$) behavior in Sr$_2$FeMoO$_6$, recently termed as spin-valve type MR (SVMR), presents several anomalies that are little understood so far. The difficulty in probing the origin of this phenomenon, arising from the magnetic property of only a small volume fraction of the ferromagnetic bulk, is circumvented in the present study by the use of {\it ac} susceptibility measurements that are sensitive to the slope rather than the magnitude of the magnetization. The present study unravels a spin-glass (SG) like surface layer around each soft ferromagnetic (FM) grain of Sr$_2$FeMoO$_6$. It is also observed that there is a very strong exchange coupling between the two, generating `exchange bias' effect, which consequently creates the `valve', responsible for the unusual $MR$ effects.}
\end{abstract}

\maketitle

PACS number(s): 75.47.-m, 73.40.Gk, 72.80.Ga

\section{Introduction}

Understanding the spin-dependent tunneling of conduction electrons between two ferromagnetic metallic electrodes through an insulating barrier has been a key issue with specific applications in the field of spintronics~\cite{Wolf_science}. Such tunneling process gets modified by the application of external magnetic fields ($H$), giving rise to the well known tunneling magnetoresistance (TMR) effect. Other than fabricated multilayers, significant TMR effect has also been observed in polycrystalline ferro-/ferrimagnetic (FM/FiM) materials in which the insulating grain boundaries act as tunnel barriers. The theoretical model of TMR predicts~\cite{TMR_reference} that the peak of the resistivity (will be called $H_{c[MR]}$ from here on) in a resistance ({\it R}) {\it vs.} {\it H} measurement is expected to appear exactly at the zero net magnetization point {\it i.e.} the magnetic coercivity ($H_{c[M]}$). This prediction was duly confirmed by experiments on all major TMR materials, such as Ca-doped LaMnO$_3$~\cite{lcmo}, CrO$_2$~\cite{cro2}, Fe$_3$O$_4$~\cite{fe3o4}, or Co-Cu alloy~\cite{cocu} in all possible material forms {\it e.g.} annealed or cold-pressed polycrystals or polycrystalline films. However, drastic exceptions have recently been observed~\cite{deteresa_fere_PRB,MR_prl,our_jpcm} in a family of double perovskite TMR materials~\cite{koba,ssc,prl,serrate_review}, especially in their cold pressed form~\cite{MR_prl,our_jpcm}, where $H_{c[MR]}$ has been found to be several times larger than $H_{c[M]}$. This highly anomalous observation was rationalized, for the representative case of Sr$_2$FeMoO$_6$ (SFMO), on the basis of a hypothesis that the insulating barriers between the soft ferromagnetic, metallic grains are composed of relatively hard magnetic material and the phenomenon was termed as SVMR~\cite{MR_prl}. Surprisingly, there has been no attempt to test the above mentioned hypothesis so far, in spite of its obvious importance to validate a new class of $MR$ materials, possessing an unusual switching behavior. This is presumably driven by the extreme difficulty, if not the impossibility, to separately identify the magnetic signal of the boundary material by any standard measurement, in the background of the overwhelmingly large contribution from the bulk.

In order to probe the true nature of these barriers and the origin of this newly discovered SVMR phenomenon, we carried out careful {\it ac} susceptibility measurement on a highly ordered polycrystalline sample of Sr$_2$FeMoO$_6$, followed by specifically designed magnetization and magnetoresistance measurements. These experiments could separate the barrier layer signal from the bulk FM signal due to their significantly different frequency responses to the applied $ac$ magnetic field. Most importantly, it was observed that the origin of SVMR is distinctly different from earlier speculation~\cite{MR_prl}. For example, our results conclusively prove a spin-glass (SG) behavior at the grain surfaces instead of the previously anticipated hard ferromagnetism. Most importantly, the present study reveals that SVMR arises due to the pinned ferromagnetic spins at the FM/SG interface as a result of exchange coupling between the ferromagnetic core and spin-glass like surface of a grain. The decisive proof for this model came from the observation of disappearing SVMR signal above the SG transition.

\section{Experimental}

The synthesis of ordered cold-pressed pellets of Sr$_2$FeMoO$_6$ have been discussed previously~\cite{koba,ssc,MR_prl}. The excellent degree of phase purity and $B$-site Fe/Mo ordering ($\sim$90\%) were confirmed in a Bruker AXS: D8 Advance x-ray diffractometer, while the magnetic and magnetotransport measurements were carried out in a Cryogenics PPMS as well as in a Quantum Design SQUID. A part of the $MR$ measurements were also performed in the longitudinal geometry using a homemade resistivity setup equipped with a superconducting magnet from Oxford Instruments. Linear and non-linear ac-susceptibility was measured using a homemade setup~\cite{alok_ac}. The XPS microscopy experiments were carried out with
the SPEM at the spectroscopy for chemical analysis (ESCA) microscopy beamline at the Elettra synchrotron facility in Trieste. In SPEM the incident photon beam is focused onto the sample to a spot of diameter, D, smaller than 0.15 $\mu$m using zone plate optics. The imaging mode in
SPEM maps the lateral distribution of elements by collecting photoelectrons with a selected energy while scanning the specimen with respect to the
focused beam.~\cite{luca1,luca2}.

\section{Results and Discussion}

In Fig. 1(a), one scanning electron microscopy (SEM) image from a typical cold pressed SFMO pellet is shown, which exhibits the sample morphology, grain sizes (3-5 $\mu$m) and the average grain size dispersion ($\pm$30\%). The grain sizes confirm that our sample is far away from the single domain regime of a ferromagnet. The linear {\it ac} susceptibility (${\chi}_1^\prime$) as a function of temperature with different frequencies, at an applied {\it ac} field of 2.33 Oe, are shown in panel (b). The experimental curves have been stacked arbitrarily along the $y$-axis to achieve better visual clarity. Clearly, a broad peak is observed around 190 K, which shifts towards lower temperatures with decreasing frequency (shown by the arrow). The imaginary part of the linear susceptibility (${\chi}_1^{\prime\prime}$) {\it vs.} $T$ (see inset to Fig. 1(b)) reveals the frequency dispersion of the signal more distinctly, albeit at a lower temperature of around 160 K. It is rather a common feature that the peak in ${\chi}_1^{\prime\prime}$ appears at a lower temperature than the same in the ${\chi}_1^\prime$, and has been observed in many well-known systems~\cite{LCaNiO_PRB,LSMCrO_PRL,YBFO_CM}, but the temperature difference here ($\approx$~30~K) seems to be unusually large. However, even such behavior has been observed earlier~\cite{alloy_PRB}, especially in systems with coexisting ferromagnetic and spin-glass features, appearing from the grains and the disordered grain boundaries, respectively~\cite{Fe_PRL}. It is to be noted that the ferromagnetic $T_c$ of Sr$_2$FeMoO$_6$ lies well above 400 K and therefore, this low temperature transition can only be associated with a coexisting metastability in the sample, arising from a very different magnetic origin. Now, a frequency dependence in ${\chi}_1$ is often considered sufficient to assign a spin-glass phase in many systems but it is also known for many decades that many long-range ordered systems or superparamagnets exhibit similar bulk susceptibility behavior, although they might arise from completely different physical origins~\cite{ac_chi,alok_book}. Therefore, mere observation of frequency dependence in ${\chi}_1$ is not conclusive enough for determining the exact magnetic ground state, especially for systems which are not exactly tailor made canonical spin-glasses. It is here that the measurement and analysis of higher order susceptibility become crucial as the higher order susceptibilities can provide the much needed information to differentiate among the variety of magnetic orders hidden in the system, {\it e.g.} non-linear susceptibility was introduced as a direct probe for the divergence of Edwards-Anderson order parameter, signifying the onset of a spin-glass transition~\cite{rmp}. In a pioneering work, this issue has been taken up for Au$_{96}$Fe$_4$, known to be a spin-glass, and Co$_{97}$Co$_3$, which is a conventional superparamagnet~\cite{bitoh_jpsj}, while many such examples can consequently be drawn~\cite{Alok, AKM}.

In general, the nonlinearity of magnetization in the presence of a magnetic field is given by~\cite{chi_expan} the series expansion,

\begin{equation}
m = m_0 + \chi_1 h + \chi_2 h^2 + \chi_3 h^3 + ...
\end{equation}

\noindent where $m_0$ is the spontaneous magnetization, $h$ is the applied alternating field, and $\chi_1$ (linear), $\chi_2$, $\chi_3$ are first-, second-, and third order susceptibilities. It is well known that only for a true spin-glass system, the real part of the third order susceptibility ($\chi_3^\prime$) increases below the glass transition temperature ($T_g$) as the amplitude of the measuring {\it ac} field tends to zero ~\cite{rmp,bitoh_jpsj,chi3}. In the inset to Fig. 1(c), the $|\chi_3^\prime|$ {\it vs.} field curves at different temperatures are shown. It is evident that $|\chi_3^\prime|$ exhibits glassy dynamics already at 183 K, consistent with the peak observed in $\chi_1^\prime$($T$) curves, while the same at 250 K displays behavior corresponding to the expected Sr$_2$FeMoO$_6$ FM state. This fact becomes even more evident from the log-log plot of the $|\chi_3^\prime|~~vs.~~T$ data (main panel of Fig. 1(c)), where linearity is observed for the data taken at $T$ = 183 K, indicating clear divergence in non-linear susceptibility - an unique characteristics of a spin-glass. Therefore, these critical experiments conclusively establish the presence of a minor spin-glass component in our sample at $T \le$ 183 K, and consequently, it can be concluded that the broad peak in the $|\chi_1^\prime|~~vs.~~T$ data indicates the onset of a true spin-glass transition. However, as the observed {\it ac} signal is composed of a magnetic metastability as well as a predominant long range FM signal, it is a nontrivial task to accurately determine the $T_g$ of the SG phase, and moreover the broadness of the transition indicates a wide distribution of exchange interactions in the system, negating the possibility of any uniquely defined spin-glass ordering temperature for the minority phase.

It is important to note here that SVMR effect arises due to the magnetic interference of the insulating tunnel barrier to the TMR process. Obviously, in case of these cold pressed samples only the magnetically active grain surfaces could be identified as this spurious tunnel barrier. However, still to establish the insulating nature of the grain surface, we have carried out photoemission microscopy experiments on our sample. At first, an image of a particular sample region was created using the imaging spectromicroscopy, and then several energy distribution curves were collected from different selected areas of this surface, illuminated by the focused beam ($\mu$-spot spectroscopy)~\cite{luca1,luca2}. In Fig. 2(a), two representative wide range spectra below the Fermi energy, collected from the deep inside and the surface region of the grains, are shown. The large shift in the spectrum from the boundary region towards higher binding energy, arising due to possible `charging effect', immediately confirms the significant insulating nature of the surface phase. Thereafter, we attempted to establish the magnetic identity of this insulating surface component since that it needed to correctly understand SVMR. Although it might be anticipated that the observed SG signal is associated with this unknown surface phase~\cite{serrate_prb,g-fe2o3_PRl,fe3o4-press_PRB}, such an assumption requires experimental confirmation. Notably, if this hypothesis is indeed valid, every grain should possess a FM/SG interface, and as a result unusual effects like `exchange bias' might be observed~\cite{g-fe2o3_PRl,fe3o4-press_PRB}, provided the SG layer has large anisotropy ~\cite{ferro-sg}. However, no clear exchange bias effect was observed in our samples by standard field-cooled (FC) and zero-field-cooled (ZFC) $M(H)$ measurements, probably because the grains are of $\mu$m sizes (Fig. 1) and the small contribution from the pinned ferromagnetic spins to the total magnetization, if any, is buried under the the large undistorted core signal~\cite{lcmo_rams,nio_prl}. However, as the intergrain tunneling process critically depends on the alignment of spins near the surface of the grains (tunnel barrier), it could be expected that the effect of exchange pinning might be observed in the $MR$($H$) measurements. Accordingly, we investigated the FC and ZFC $MR(H)$ in search of possible exchange bias effects. In Fig. 2(b), we show an expanded view of the FC and ZFC $MR(H)$ curves at 5~K. Distinct shifts (indicated by brown, red and blue arrows) of the FC $MR$ peaks ($H_{c[MR]}$) with respect to the symmetric ZFC loop along the field axis are found, while the shifts changed signs depending on the direction of the applied fields during cooling. Here, a clear analogy can be drawn with the asymmetric shift of the FC $M(H)$ hysteresis loops, which is the key point of all exchange bias systems~\cite{nogues_PR}. Such shifts in the field sweep measurements are known to arise due to the pinned ferromagnetic spins at the interface, originated as a consequence of strong exchange coupling between the soft ferromagnetic and hard antiferromagnetic/ferrimagnetic or spin-glass layers~\cite{nogues_PR}. Therefore, it is possible to draw two important conclusions here. Firstly the spurious SG signal must be associated with the surface phase {\it i.e.} the tunnel barrier, giving rise to definite FM/SG exchange coupling across the core/surface interfaces in each grain, pinning the ferromagnetic spins near the interface, and producing the asymmetric shifts in the $MR(H)$ curves. Recently, we have shown that nanosized Sr$_2$FeMoO$_6$ particles, with largely enhanced surface volume, indeed exhibit clear exchange bias effect in the $M$($H$) measurements~\cite{jap}, which further supports this conclusion. Secondly, as the field at which the $MR$ peaks is susceptible to cooling history and exchange biasing, it could be understood that the TMR response is critically following the field response of the pinned ferromagnetic spins, and contrary to earlier proposal~\cite{MR_prl} it has to be these spins which must be acting as the `valve', giving rise to the SVMR response. It is obvious that these pinned FM spins at the interface are hard to rotate and can effectively act as scattering centers for the softer core spins that are trying to tunnel through the insulating barrier. To confirm the reliability of the observed exchange bias, we show the ZFC $M(H)$ data from our sample n Fig. 2(c), which comes out of irreversibility at a field of only 0.1 Tesla, while the ZFC $MR(H)$ measurements were carried out within a field range of -0.5 to +0.5 Tesla (Fig. 2(d)). This observation testifies that the observed loop shift is not arising due to the {\it so called} minor loop effect~\cite{minor_loop}. Another experimental aspect that is worth noting here is the observed significant asymmetries between the highest resistivity values in the positive and negative field quadrants in any particular FC $MR(H)$ measurement, which are also related to exchange biasing (Fig. 2b).

It is to be noted that exchange bias is known to disappear if the maximum applied field ($H_{max}$) in a field sweeping measurement becomes too high~\cite{eb_standard}, that is if the applied field becomes large enough to orient even the hard ferrimagnetic or SG layers. Therefore, $H_E$ is expected to fall with increasing $H_{max}$, which is exactly what is observed in this case, shown in Fig. 3(a). In Fig. 3(b), we show the dependence of $H_E$ on the cooling field. This behavior is reminiscent of all exchange bias systems, FM-SG bilayer being one of them~\cite{fe3o4-press_PRB,FM-SG}.

In order to check all the hypothesis made so far, we also attempted to carry out extensive numerical simulations, where the model considered is a modification of that used in our earlier works~\cite{MR_prl}. A schematic representation of the present model and the magnitude of different interactions are shown in the left panel of Fig. 4. Here, there is a magnetic interfacial coupling between the hard SG layer `h' and the surface of the soft FM grains. This coupling gives rise to a `pinned' layer `p' at the surface of the soft FM grains, crucial for the exchange bias effect. The hard SG layer has both ferro ($J_{FM}(p-h)$) and antiferro ($J_{FM}(p-h)$) couplings with the pinned layer. There is also a weaker, ferro coupling between the soft layer and the pinned layer ($J_{FM}(s-p)$), arising essentially from the remnant ferromagnetism at the grain surface. There are also randomly oriented anisotropy axes within the spin glass layer, designed to make it hard, without giving any preferential
axis direction. Since this pinned layer is at the core/surface interface of the grain, it controls the switching of the tunneling, and the exchange bias effect is reflected in the $MR$($H$) curves. We have changed the method of calculating the Landauer conductance to Non-Equilibrium Green's Function (NEGF)  method instead of the Transfer Matrix method used in the earlier work~\cite{MR_prl} as this is more stable and allows larger sizes. The response of this system in absence of a bias field is identical to that of our previous model, however, in presence of a bias cooling field, the response is very different. The results of the simulation is presented in the right panel of Fig. 4. Evidently, the experimental observations of the shifts along the field axis as well as the $MR$ axis in the $MR$($H$) curves are captured very well in these simulations. It is to be noted that we considered the hard layer as electronically transparent, {\it i.e.}, the spins in this layer do not affect the transport, otherwise the observed $MR$ would have traced out a minor loop, which it does not. In ref. 30, a model similar to ours had been proposed, where the defective surface with random anisotropy is similar to the hard SG layer proposed here. Since we have actually done a full-fledged simulation of this model on a finite size system involving Monte Carlo and NEGF, additionally we have been able to probe the pinned layer, the exchange bias effect as well as the hysteresis, all of which are non-equilibrium phenomena.

However, the decisive point for the whole understanding is the assumed direct connection between the appearance of the SG layer at the surface, generating the {\it valve} through exchange coupling, and the consequent arrival of SVMR. Therefore, the acid test for such an understanding would be to study the temperature dependence of SVMR, which should disappear above $T_g$ of the glassy skin layer as a result of extinguished pinning effect at the interface. In Fig. 5(a), we show $M$ $vs.$ $H$ behavior of our sample at few different temperatures. Following the expected line, the saturation moment as well as the coercivity (see inset) is found to decrease regularly with increasing temperature. Similarly, in Fig. 5(b), $MR$($H$) behaviors at same temperatures and their resistivity peaks (inset) are shown. Next, we have extracted the temperature variation of SVMR from these experiments and the results are summarized in Fig. 6.

In Fig. 6(a), variations of $H_{c[M]}$ and $H_{c[MR]}$ as a function of temperature is shown. Expectedly, the separation between the two quantities, {\it i.e.} the SVMR effect, is most severe at the lowest temperature. But most interestingly, the distinction between the two is found to diminish gradually with increasing temperature and disappear around 200 K, {\it i.e.} at the estimated $T_g$ of the SG-like skin (see Fig. 1). We have also observed that the exchange bias in $MR$($H$) completely vanishes above the glass transition temperature, confirming the close connection between the setting up of the spin-glass order at the grain surface, the exchange bias, and the SVMR. In the inset to Fig. 6(a), \%$MR$ is plotted against $(M/M_s)^2$ at 250~K ({\it i.e.} above the $T_g$), which follows the predicted straight line like behavior at the low field region, where the intergrain TMR is known to be most pronounced. The deviation from the linear behavior at higher fields might appear due to intragrain tunneling mechanism, often discussed for Sr$_2$FeMoO$_6$~\cite{mar_prl}. Fig. 6(b) shows the development of $MR$ as a function of the bulk $M$, from which the absence of SVMR above $T_g$ is understood more clearly as the $MR$ is found to touch the zero value ($H_{c[MR]}$) exactly at the zero magnetization point ($H_{c[M]}$) point at 250~K. This behavior is consistent with any other conventional TMR material~\cite{lcmo,cro2,fe3o4,cocu}. However, at 5~K the tunneling needed much higher field to kickoff, giving rise to SVMR.

\section{Conclusions}

In summary, our detailed experiments clearly establish the presence of a spin-glass like phase at the surface of each grain, which is spontaneously formed during the mechanical creation of the small grain powders of Sr$_2$FeMoO$_6$. However, this SG layer immediately exchange couples with the core ferromagnetic spins through the interface and produces the `valve' that stops the tunneling of conduction electrons between two adjacent grains even though the core of each grain is largely aligned in the same direction as the applied magnetic field. This phenomena gives rise to the unprecedent effect of the SVMR, observed in double perovskites in general~\cite{deteresa_fere_PRB,MR_prl,our_jpcm}. Understandably, this effect can persist only till the SG phase exist and immediately gets perished as soon as the experimental temperature goes above the $T_g$.

\acknowledgments

SR and DDS thank the DST and BRNS, India for financial support. SM and SJ thank CSIR, India for fellowship. PS thanks Dr. A. Agarwal for useful discussions.

\newpage

\begin{figure}
\begin{center}
\resizebox{8cm}{!}
{\includegraphics*[220pt,322pt][462pt,615pt]{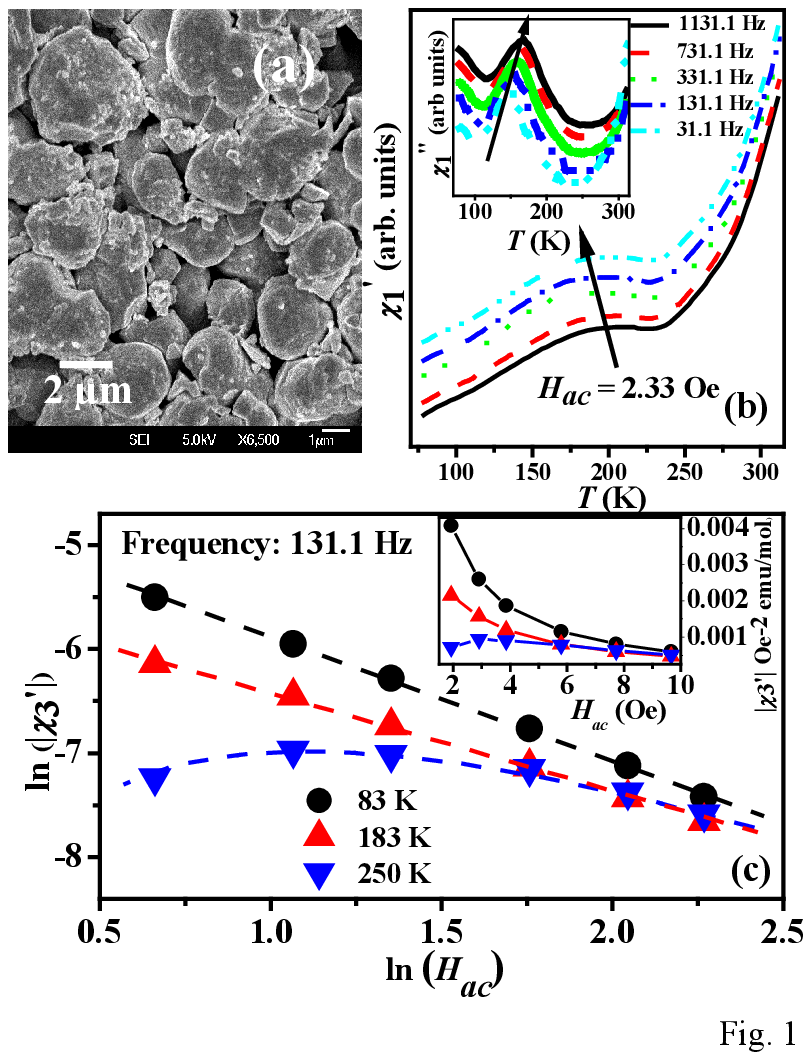}} \\
\caption{(color online) (a) shows a SEM image of the cold pressed sample. (b) Real and imaginary (inset) parts of linear $ac$ susceptibility {\it vs.} $T$. (c) the log-log plot of $|\chi_3^\prime|$ {\it vs.} $H$, while the linear data are shown in the inset.}
\end{center}
\end{figure}

\begin{figure}
\begin{center}
\resizebox{8cm}{!}
{\includegraphics*[20pt,65pt][540pt,825pt]{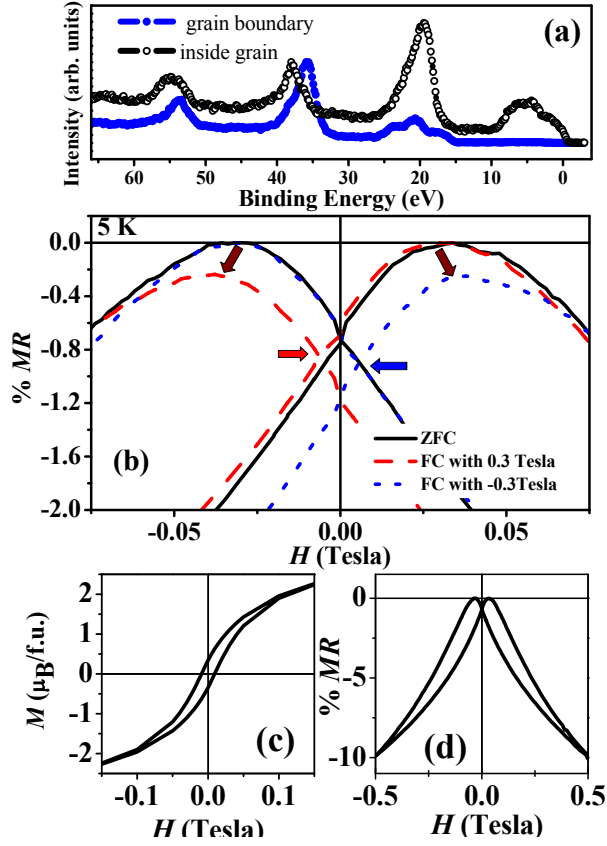}} \\
%%\vspace{-1 in}
\caption{(color online) (a) Two representative wide range, near valence band spectra from the sample. (b) Expanded $MR(H)$ data, while the full field range data is shown in panel (d). Panel (c) shows $M(H)$ of the same sample. All the measurements have been carried out at 5~K.}
\end{center}
\end{figure}

\begin{figure}
\begin{center}
\resizebox{8cm}{!}
{\includegraphics*[7pt,435pt][560pt,720pt]{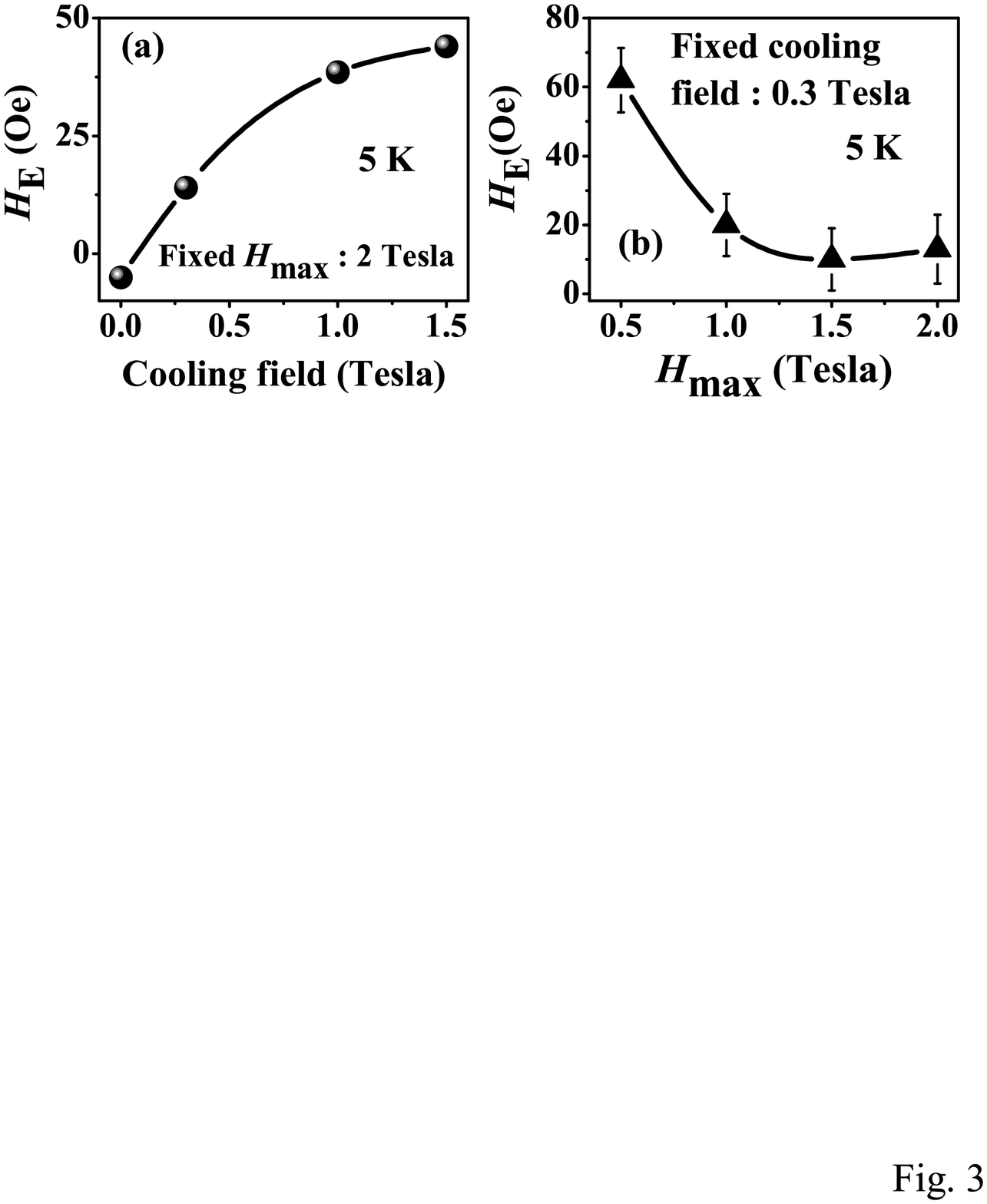}} \\
%%\vspace{-1 in}
\caption{(color online) The variations of the exchange field ($H_E$) as a function of maximum measuring field in the field sweeping measurements and as a function of cooling field are shown in panels (a) and (b), respectively.}
\end{center}
\end{figure}

\begin{figure}
\begin{center}
\resizebox{7cm}{!}
{\includegraphics*[47pt,420pt][569pt,733pt]{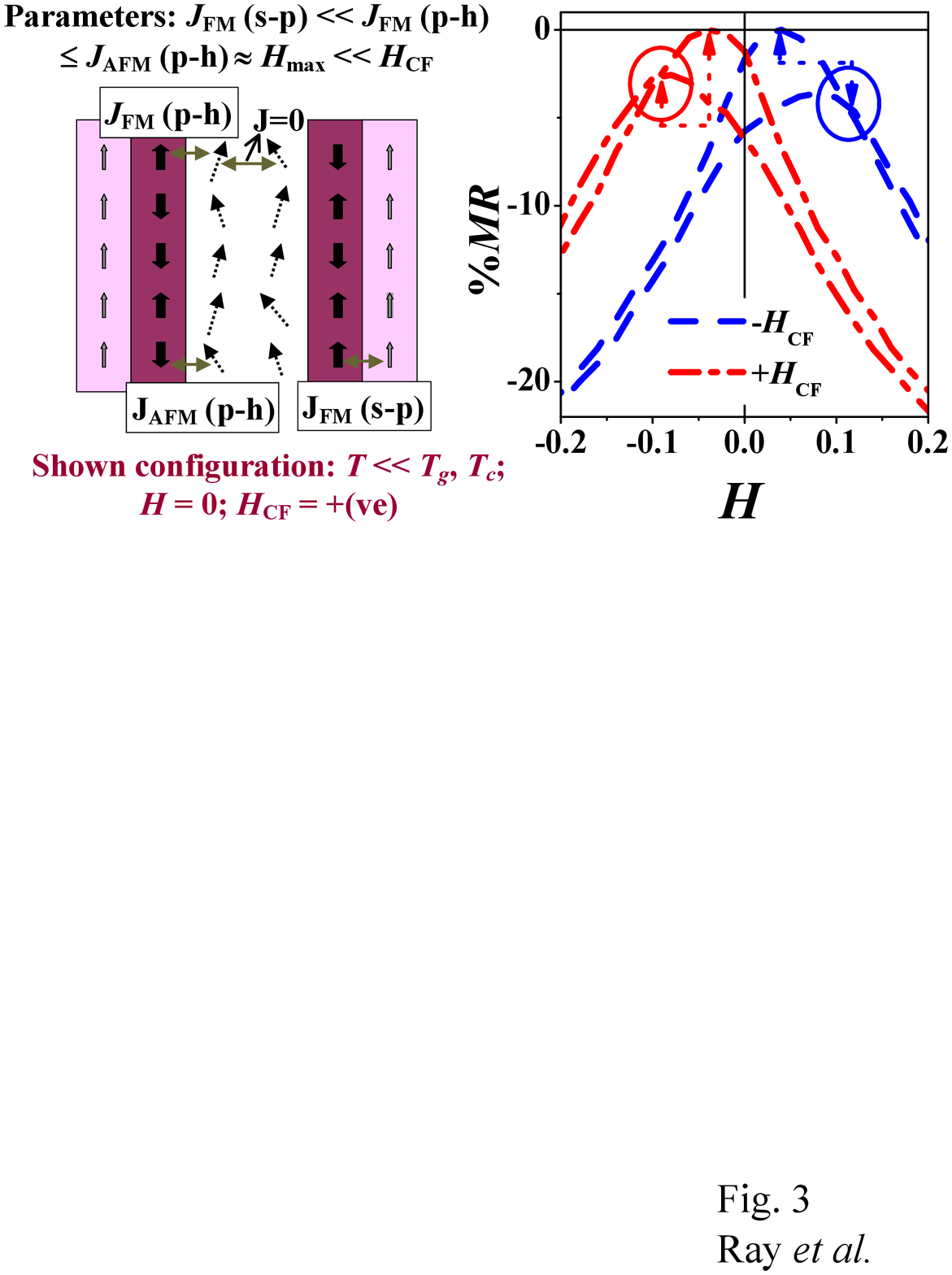}} \\
%\vspace{-1 in}
\caption{(color online) A schematic of the model is shown on the left half, while the simulated MR data is shown on the right half in an expanded scale. In the schematic, Light pink: Soft Layer (s); Purple : Pinned layer (p); White: Hard layer (h); Vertical/canted arrows symbolize
the spins, while horizontal double headed arrows show the coupling between the different layers: $J_{FM}(s-p)$ is the ferromagnetic coupling between `s' and `p' layers. Some spins of rough interface between the `p' and the `h' layers have ferro coupling $J_{FM}(p-h)$, while
others have antiferro coupling $J_{AFM}(p-h)$.  $H_{max}$ is the maximum value of the sweeping field, while $H_{CF}$ is the bias field.}
\end{center}
\end{figure}

\begin{figure}
\begin{center}
\resizebox{8cm}{!}
{\includegraphics*[32pt,37pt][545pt,815pt]{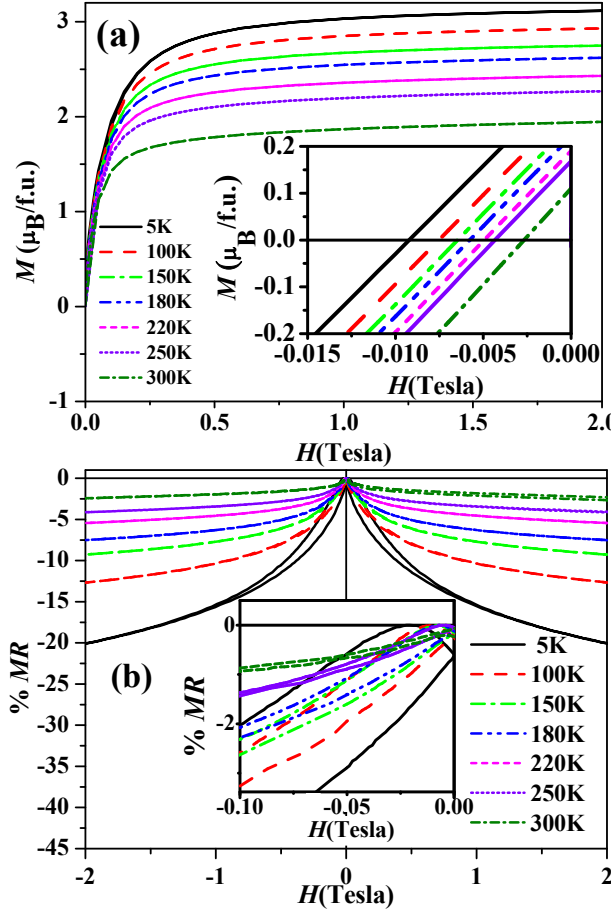}} \\
%%\vspace{-1 in}
\caption{(color online) (a) $M$($H$) behaviors at different temperatures are shown while the inset shows the coercivity variation in an expanded scale. (b) similarly, $MR$($H$) behaviors at different temperatures are exhibited while the inset shows the resistivity peak variation in an expanded scale.}
\end{center}
\end{figure}

\begin{figure}
\begin{center}
\resizebox{8cm}{!}
{\includegraphics*[2pt,397pt][570pt,769pt]{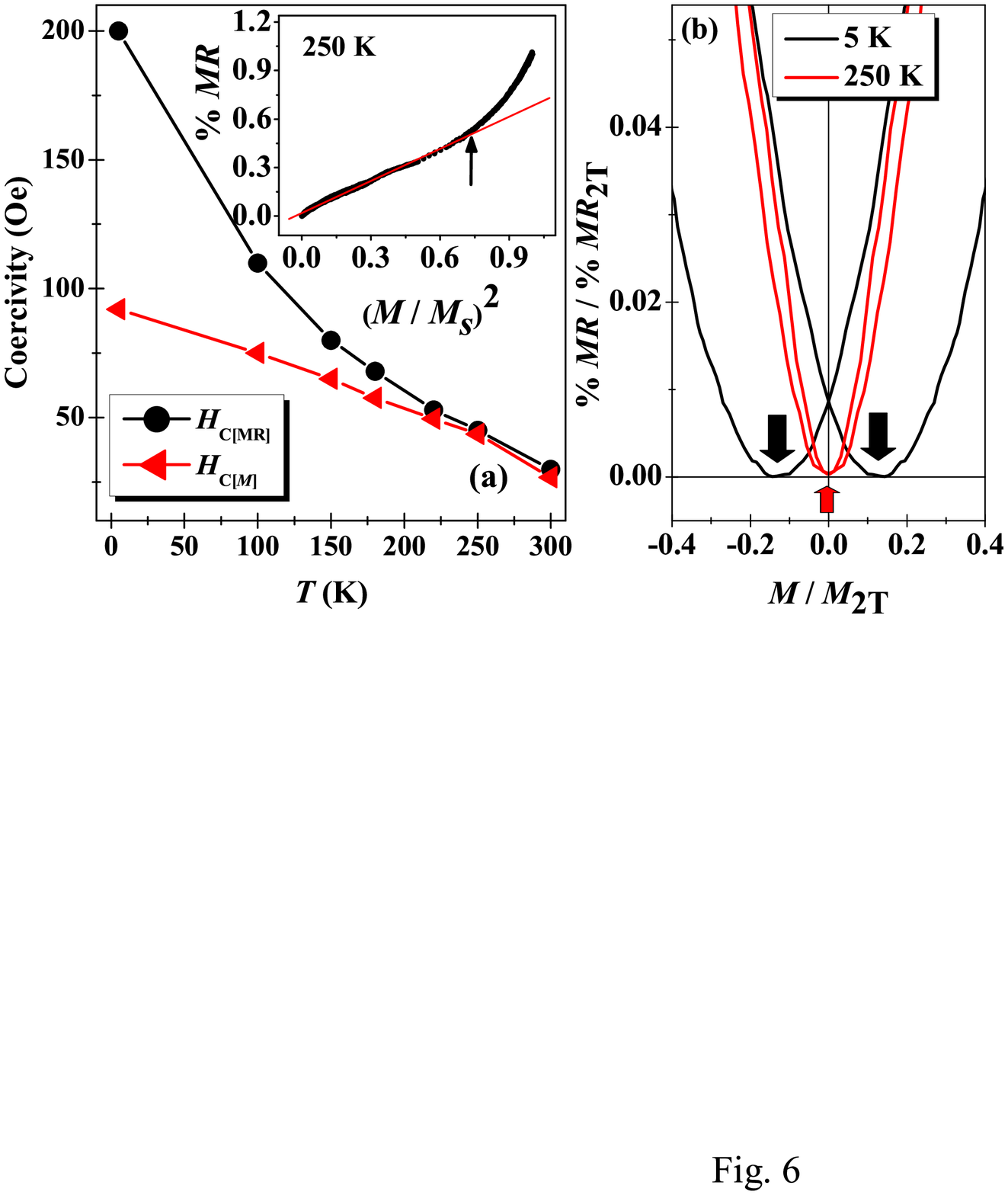}} \\
%\vspace{-1 in}
\caption{(color online)The temperature variation of $H_{c[M]}$ and $H_{c[MR]}$ are shown in panel (a). The inset to panel (a) shows the near linear dependence of $MR$ on $M^2$ at 250~K within at least the low-field region. The $MR$ {\it vs.} $M$ dependence at 5~K and 250~K are shown in (c).}
\end{center}
\end{figure}

\end{document}